\documentstyle[12pt]{article}
\advance \topmargin by -\headheight
\advance \topmargin by -\headsep
\evensidemargin \oddsidemargin
\begin{document}
\begin{center}
{\bf Dominance of a Dynamical Measure 
\\
and Disappearance of the Cosmological Constant}
 \end{center}

\bigskip
\begin{center}

\bigskip

E.I.Guendelman\footnote{Electronic address: GUENDEL@BGUmail.BGU.AC.IL} and
        A.B.Kaganovich\footnote{Electronic address: ALEXK@BGUmail.BGU.AC.IL}

{\it Physics Department, Ben Gurion University, Beer
Sheva, Israel}
\end{center}

\bigskip

\begin{abstract}

We consider an action which consists of two terms: the first            
$S_{1}=\int L_{1}\Phi d^{4}x$
 and the second $S_{2}=\int L_{2}\sqrt{-g}d^{4}x$
where $\Phi$ is a measure which has to be determined dynamically. $S_{1}$ 
satisfies the requirement that the transformation $L_{1}\rightarrow 
L_{1}+const.$ does not effect equations of motion. In the first order 
formalism, a constraint appears which allows to solve $\chi =
\Phi/\sqrt{-g}$. Then, in a true vacuum state (TVS), $\chi\rightarrow\infty$
and in the conformal Einstein frame no singularities are present, the 
energy density of TVS is zero without fine tuning of any scalar potential 
in $S_{1}$ or $S_{2}$. When considering only a linear potential for a 
scalar field $\phi$ in $S_{1}$, the continuous symmetry 
$\phi\rightarrow\phi+const$ is respected. Surprisingly, in this case SSB 
takes place while no massless ("Goldstone") boson appears. 

 \end{abstract}

\pagebreak

{\bf 1. Introduction.} 

\bigskip

One of the greatest puzzles of modern physics is 
the Cosmological Constant Problem (CCP) which consists of the fact that 
present day vacuum energy is zero or so small in Planck units.

In a series of papers \cite{GK1}-\cite{GK4} we have developed the so called 
Nongravitating 
Vacuum Energy (NGVE) theory, where the action of the theory is assumed to be
\begin{equation}
S=\int\Phi Ld^{4}x
\label{1}
\end{equation}
$\Phi$ being a total derivative of something and also a scalar density, 
as for example can be achieved by constructing it out of four scalar 
fields $\varphi_{a}\quad (a=1,...,4)$:
\begin{equation}
\Phi \equiv \varepsilon_{abcd}
\varepsilon^{\mu\nu\alpha\beta}
(\partial_{\mu}\varphi_{a})
(\partial_{\nu}\varphi_{b})
(\partial_{\alpha}\varphi_{c})
(\partial_{\beta}\varphi_{d})
\label{2}
\end{equation}
In this case $L$ can be changed as in $L\rightarrow L+const$ without 
affecting the equations of motion. Notice that $\Phi$ is a measure 
independent of $g_{\mu\nu}$ as opposed to the case of GR where the 
measure of integration is $\sqrt{-g}$. In what follows we will call 
$\Phi$ the {\em Dynamical Measure} (DM) since its value is dynamically 
determined in terms of all the fields of the theory through the equations 
of motion as we will see. 

In Refs. \cite{GK2}-\cite{GK4} we have considered 
\begin{equation}
L=-\frac{1}{\kappa}R(\Gamma,g)+L_{m}
\label{3}
\end{equation} 
where the matter Lagrangian density $L_{m}$ does not contain 
$\varphi_{a}$ dependence and $R(\Gamma,g)$ is the scalar  curvature
$R(\Gamma,g)=g^{\mu\nu}R_{\mu\nu}(\Gamma)$ of the space-time of the
affine connection  $\Gamma^{\mu}_{\alpha\beta}$: \,
$R_{\mu\nu}(\Gamma)=R^{\alpha}_{\mu\nu\alpha}(\Gamma)$, \,
$R^{\lambda}_{\mu\nu\sigma}(\Gamma)\equiv
\Gamma^{\lambda}_{\mu\nu,\sigma}-\Gamma^{\lambda}_{\mu\sigma,\nu}+
\Gamma^{\lambda}_{\alpha\sigma}\Gamma^{\alpha}_{\mu\nu}-
\Gamma^{\lambda}_{\alpha\nu}\Gamma^{\alpha}_{\mu\sigma}$.
If in addition to  this we assume that $L_{m}$ contains 4-index field 
strength and also scalar fields with generic potentials with no special 
fine tuning imposed on them, we obtain generically \cite{GK4}
that Eqs. (\ref{1})-(\ref{3}) solve the cosmological constant problem.

Here we will see that it is possible to extend considerably the class of 
theories where the CCP is solved in the same spirit. In fact we will see 
that it is possible to add an {\em explicit} cosmological constant term 
to Eq.(\ref{1}), or even more general, a coupling of a scalar field   
$\phi$ to $\sqrt{-g}$ of the form
\begin{equation}
\int U(\phi)\sqrt{-g}d^{4}x
\label{4}
\end{equation}
without destroying the mechanism that drives the effective cosmological 
constant to zero which can work even in the absence of a 4-index field
strength. Further generalizations and extensions of the theory can also 
be made, without destroying the basic mechanism which solves the CCP.

\bigskip

\begin{center}
{\bf 2. Dynamical measure dominance in the true vacuum state 
\\
and solution of the CCP}
\end{center}

To demonstrate how the theory works, we start here from the simplest 
model in the first order formalism including scalar field $\phi$ and gravity
according to the prescription of the NGVE principle and in addition to 
this we include the standard cosmological constant term. Possible reasons 
for such kind of structure will be discussed at the end of this letter. 
So, we consider an action 
\begin{equation}
S=\int L_{1}\Phi d^{4}x + \int\Lambda\sqrt{-g}d^{4}x
\label{5}
\end{equation}
where 
\begin{equation}
L_{1}=-\frac{1}{\kappa}R(\Gamma,g)+\frac{1}{2}g^{\mu\nu}\phi_{,\mu}\phi_{,\nu}
-V(\phi)
\label{6}
\end{equation}

Performing the variation with respect to the measure fields $\varphi_{a}$ 
(see Eq.(2)) we obtain equations 
$A^{\mu}_{a}\partial_{\mu}L_{1}=0$
where $A^{\mu}_{b}=\varepsilon_{acdb}
\varepsilon^{\alpha\beta\gamma\mu}
(\partial_{\alpha}\varphi_{a})
(\partial_{\beta}\varphi_{c})
(\partial_{\gamma}\varphi_{d})$.
It is easy to check (see \cite{GK1}-\cite{GK3}) that if $\Phi\neq 0$, then 
it follows from the last equations that $L_{1}=M=const$

Varying the action (5) with respect to $g^{\mu\nu}$ we get
\begin{equation}
\Phi(-\frac{1}{\kappa}R_{\mu\nu}(\Gamma)+\frac{1}{2}\phi_{,\mu}\phi_{,\nu})
-\frac{1}{2}\sqrt{-g}\Lambda g_{\mu\nu}=0
\label{7}
\end{equation}

Contracting Eq.(7) with $g^{\mu\nu}$ and using equation $L_{1}=M$ we 
obtain the constraint
\begin{equation}
M+V(\phi)-\frac{2\Lambda}{\chi}=0
\label{8}
\end{equation}
where we have defined the scalar field $\chi\equiv\Phi/\sqrt{-g}$.

The scalar field $\phi$ equation is
\begin{equation}
(-g)^{-1/2}\partial_{\mu}(\sqrt{-g}g^{\mu\nu}\partial_{\nu}\phi)+
\sigma_{,\alpha}\phi^{,\alpha}+V^{\prime}=0
\label{9}
\end{equation}
where $\sigma\equiv\ln\chi$ and $V^{\prime}\equiv dV/d\phi$.

The solution of the equation obtained by variation of the connection
$\Gamma^{\alpha}_{\mu\nu}$ may be represented in the form 
$\Gamma^{\alpha}_{\mu\nu}=\{ ^{\alpha}_{\mu\nu}\}+
\Sigma^{\alpha}_{\mu\nu}(\sigma)$
where $\{ ^{\alpha}_{\mu\nu}\}$ are the Christoffel's connection 
coefficients and $\Sigma^{\alpha}_{\mu\nu}$ is a function of derivatives 
of $\sigma$. After 
making use the $\lambda$-symmetry \cite{Ein} of the action (5) 
\begin{equation}
\Gamma^{\prime\alpha}_{\mu\nu}=\Gamma^{\alpha}_{\mu\nu}+
\delta^{\alpha}_{\mu}\lambda_{,\nu}
\label{10}
\end{equation}
the antisymmetric part of $\Sigma^{\alpha}_{\mu\nu}(\sigma)$ can be 
eliminated and we get
\begin{equation}
\Sigma^{\alpha}_{\mu\nu}(\sigma)=
\frac{1}{2}(\delta_{\mu}^{\alpha}\sigma_{,\nu}+
            \delta_{\nu}^{\alpha}\sigma_{,\mu}-
            \sigma_{,\beta}g^{\alpha\beta}g_{\mu\nu})
\label{11}
\end{equation}

The derivatives of the field $\sigma$ enter both the gravitational 
equation (7) (through the connection) and in the scalar field equation
(9). By a conformal transformation 
\begin{equation}
g_{\mu\nu}\rightarrow\overline{g}_{\mu\nu}=\chi g_{\mu\nu}; \qquad 
\phi\rightarrow\phi \label{12}
\end{equation}
to an "Einstein picture" and using the constraint (8) we obtain the 
canonical form of equations for the scalar field $\phi$
\begin{equation}
(-\overline{g})^{-1/2}\partial_{\mu}(\sqrt{-\overline{g}}\quad
\overline{g}^{\mu\nu}\partial_{\nu}\phi)+V_{eff}^{\prime}(\phi)=0
\label{13}
\end{equation}
and the gravitational equations in the Riemannian space-time with metric 
$\overline{g}_{\mu\nu}$
\begin{equation}
R_{\mu\nu}(\overline{g}_{\alpha\beta})-
\frac{1}{2}\overline{g}_{\mu\nu}R(\overline{g}_{\alpha\beta})
=\frac{\kappa}{2}T^{eff}_{\mu\nu}(\phi)
\label{14}
\end{equation}
where $T^{eff}_{\mu\nu}(\phi)=\phi_{,\mu}\phi_{,\nu}-
\frac{1}{2}\overline{g}_{\mu\nu}
\phi_{,\alpha}\phi_{,\beta}\overline{g}^{\alpha\beta}+
V_{eff}(\phi)\overline{g}_{\mu\nu}$,
\begin{equation}
V_{eff}(\phi)=\frac{1}{4\Lambda}[M+V(\phi)]^{2}
\label{15}
\end{equation}
and 
\begin{equation}
V_{eff}^{\prime}(\phi)=\frac{1}{2\Lambda}[M+V(\phi)]V^{\prime}(\phi)
\label{16}
\end{equation}

We see that for any analytic function $V(\phi)$, the effective potential 
in the Einstein picture has an extremum, i.e. $V_{eff}^{\prime}=0$, 
either when $V^{\prime}=0$ or $V+M=0$. The extremum $\phi=\phi_{1}$ where 
$V^{\prime}(\phi_{1})=0$ has nonzero energy density 
$[M+V(\phi_{1})]^{2}/4\Lambda$ if a fine tuning is not assumed. In 
contrast to this, if $\Lambda >0$, the state $\phi =\phi_{0}$ where 
$V(\phi_{0})+M=0$ is the absolute minimum and therefore $\phi_{0}$ is a 
true vacuum with zero cosmological constant without any fine tuning. A mass 
square of the scalar field describing small fluctuations around 
$\phi_{0}$ is
\begin{equation}
m^{2}=\frac{1}{2\Lambda}[V^{\prime}(\phi_{0})]^{2}
\label{17}
\end{equation}  
 Let us now consider a couple of models for $V(\phi)$ and some 
interesting effects associated with them.

\bigskip

\begin{center}

{\bf 3. Model with continuous symmetry related to the NGVE principle 
\\
and SSB 
without generating a massless scalar field}

\end{center}
 
\bigskip

It is interesting to see what happens in the model for the choice 
$V=J\phi$, where $J$ is some constant. Then the action (5), (6) is 
invariant (up to the integral of total divergence) under the shift 
$\phi\rightarrow\phi +const$ which is in fact the symmetry 
$V\rightarrow V+const$ related to the NGVE principle. Notice that if we 
 consider 
the model with complex scalar field $\psi$, where $\phi$ is the phase of 
$\psi$, then the symmetry $\phi\rightarrow\phi +const$ would be the 
$U(1)$ - symmetry.

The effective potential (15) in such a model has the form 
\begin{equation}
V_{eff}=\frac{1}{2}m^{2}(\phi -\phi_{0})^{2}
\label{18}
\end{equation}
 where $\phi_{0}=-M/J$
and $m^{2}=J^{2}/2\Lambda$. We see that the symmetry 
$\phi\rightarrow\phi +const$ is spontaneously broken and mass generation 
is obtained. However, {\em no massless scalar field results from the 
process of SSB in this case, i.e. Goldstone theorem does not apply here}.

This seems to be a special feature of the NGVE - theory which allows:
1) To start with linear potential $J\phi$ without destroying the shift 
symmetry $\phi\rightarrow\phi +const$, present in the 
$\partial_{\mu}\phi \partial^{\mu}\phi$ piece, due to the coupling to the 
dynamical measure (2). This shift symmetry is now a symmetry of the 
action up to a total divergence.
2) This potential gives rise to an effective potential 
$(M+J\phi)^{2}/4\Lambda$. The constant of integration $M$ being 
responsible for the SSB.

Similar effect can be obtained even in the pure NGVE - theory as in 
Eq.(1) (without introduction of an explicit $\Lambda$-term) but with the 
use of 4-index field strength condensate. The possibility of constructing 
spontaneously broken $U(1)$ models which do not lead to associated 
Goldstone bosons is of course of significant physical relevance. One may 
recall for example the famous $U(1)$ problem in QCD \cite{U1QCD}. Also 
the possibility of mass generation for axions is of considerable 
interest. These 
issues will be developed further in elsewhere \cite{GKprep}.  
 
\bigskip

{\bf 4. Selfinteraction and SSB from quadratic scalar field potential}

\bigskip

If we choose the quadratic potential $V(\phi)=\frac{\mu^{2}}{2}\phi^{2}$,
which in the usual theory is associated with free field theory in curved 
space-time, we obtain here a very different result, i.e. that the 
effective potential in the Einstein picture is
\begin{equation}
V_{eff}=\frac{\mu^{4}}{16\Lambda}(\phi^{2}-\phi_{0}^{2})^{2}
\label{19}
\end{equation}
where $\phi_{0}^{2}=-2M/\mu^{2}$

We see that if $\Lambda >0$, spontaneous breaking of the discrete symmetry
$\phi\rightarrow -\phi$ takes place if $M\mu^{2}<0$. Yet, the vacuum energy 
at the absolute minimum $\phi =\pm |\phi_{0}|$ is identically zero. 
Furthermore, the mass of the scalar field is $m^{2}=
\frac{\mu^{4}}{2\Lambda}\phi_{0}^{2}=-\frac{M\mu^{2}}{\Lambda}$ which as 
we see depends on the integration constant $M$. The mass $m$ is therefore 
a "floating" physical parameter, since $M$ does not appear in the original 
Lagrangian but it is determined by initial conditions of the Universe. 

If $\phi$ is replaced by a complex field $\psi$ and $\phi^{2}$ by 
$\psi^{\ast}\psi$ we obtain the SSB of a continuous symmetry with 
standard consequences (as opposed to example of Sec.3). In addition, a 
model of cosmology that can include an inflationary phase taking place in 
a false vacuum and transition to a zero cosmological constant phase is 
obtained without fine tuning.

\bigskip

{\bf 5. Discussion}

\bigskip

Further generalizations, like considering a term of the form 
$\int U(\phi)\sqrt{-g}d^{4}x$ instead of 
$\Lambda\int \sqrt{-g}d^{4}x$, the possibility of coupling of scalar 
fields to curvature, etc. do not modify the qualitative nature of the 
effects described here and they will be studied in a more detailed 
publication \cite{GKprep}. For example, even in the presence of 
$V_{1}(\phi)$ the resulting effective potential vanishes when 
$V(\phi_{0})+M=0$ and it goes as 
$V_{eff}=\frac{1}{4U(\phi_{0})}(V+M)^{2}$ in the region $V+M\sim 0$.

Some sources of the type of structure being considered here are 
suggested. First of all, as mentioned in Ref.\cite{GK1}, if we start with 
only $S_{1}=\int L_{1}\Phi d^{4}x$ as a fundamental theory, possible 
radiative corrections in the effective action are severely constrained 
due to the existence of the infinite dimensional symmetry \cite{GK1}
(up to a total divergence) which consists of the infinitesimal shift of the 
fields $\varphi_{a}$ by an 
arbitrary infinitesimal function of the total Lagrangian density $L_{1}$
 \begin{equation}
\varphi_{a}\rightarrow\varphi_{a}+\epsilon g_{a}(L_{1}), \qquad \epsilon\ll 1
\label{20}
o
\end{equation}
This symmetry prevents the appearance of terms of the form $f(\chi)\Phi$ 
in the effective action with the single possible exception of $f(\chi)=
c/\chi$ where a scalar $c$ is $\chi$ independent. This 
is because in this last  case 
the term $f(\chi)\Phi =c\sqrt{-g}$ is $\varphi_{a}$ independent. This 
possibility gives rise to the cosmological constant term  in the 
effective action as in Eq.(5) while the symmetry (20) is maintained. This 
can be generalized to possible contributions of the form 
$\int L_{2}\sqrt{-g}d^{4}x$ where $L_{2}$ is $\varphi_{a}$ independent 
function of matter fields and gravity if radiative corrections generate a 
term $f(\chi)\Phi$ with $f(\chi)=L_{2}/\chi$.

One may question also the possible origin of the measure of integration 
$\Phi$. It appears to us that an interesting possibility is that this may 
correspond to a space-filling brane \cite{Polch}, as discussed in 
Ref.\cite{GK3}. It may be noticed that if we take $L_{1}=const$ in the 
action $\int L_{1}\Phi d^{D}x$ of the fundamental theory in an arbitrary 
dimension $D$, we obtain a 
purely topological theory which has been interpreted as the nontrivial 
action for a $D-2$ brane by Zaikov \cite{Z}. In 
our case, the introduction of $L_{1}\neq const$ changes drastically the 
nature of the theory, making it dynamical rather than topological, but 
may be the geometrical interpretation (or parts of it) could be retained.

\end{document}